\documentclass{IEEEcsmag_oa}

\usepackage[colorlinks,urlcolor=blue,linkcolor=blue,citecolor=blue]{hyperref}
\expandafter\def\expandafter\UrlBreaks\expandafter{\UrlBreaks\do\/\do\*\do\-\do\~\do\'\do\"\do\-}
\usepackage{upmath}
\usepackage{tabularray}
\usepackage{caption}
\usepackage{subcaption}
\usepackage{multirow}
\usepackage{tikz}
\usetikzlibrary{positioning}
\usetikzlibrary{shapes}
\usetikzlibrary{arrows}
\usepackage{pgfplots}
\usepackage{mathtools}
\usepackage{placeins}
\usepackage{dingbat}
\usepackage{booktabs}
\usepackage{footmisc}
\usepackage{comment}
\usepackage{mdframed}
\usepackage{graphicx}
\usepackage{colortbl}
\usepackage{hhline}

\newcommand\blfootnote[1]{%
	\begingroup
	\renewcommand\thefootnote{}\footnote{#1}%
	\addtocounter{footnote}{-1}%
	\endgroup
}

\PassOptionsToPackage{table}{xcolor}

\usepackage{array}
\newcolumntype{L}[1]{>{\raggedright\let\newline\\\arraybackslash\hspace{0pt}}m{#1}}
\newcolumntype{C}[1]{>{\centering\let\newline\\\arraybackslash\hspace{0pt}}m{#1}}
\newcolumntype{R}[1]{>{\raggedleft\let\newline\\\arraybackslash\hspace{0pt}}m{#1}}

\newcommand{\enquote}[1]{``#1''}

\newcommand{\REV}[1]{#1}

\begin{document}

\title{Perceptual Hash Inversion Attacks on Image-Based Sexual Abuse Removal Tools \\\normalsize (Preprint Version*)}

\author{Sophie Hawkes}
\affil{Royal Holloway, University of London, UK}

\author{Christian Weinert}
\affil{Royal Holloway, University of London, UK}

\author{Teresa Almeida}
\affil{Umeå University, Sweden and ITI/LARSyS, Portugal}

\author{Maryam Mehrnezhad}
\affil{Royal Holloway, University of London, UK}

\markboth{Hawkes et al.}{Perceptual Hash Inversion Attacks on IBSA Removal Tools}

\begin{abstract}
We show that perceptual hashing, crucial for detecting and removing image-based sexual abuse~(IBSA) online, faces vulnerabilities from low-budget inversion attacks based on generative~AI. This jeopardizes the privacy of users, especially vulnerable groups. We advocate to implement secure hash matching in~IBSA removal tools to mitigate potentially fatal consequences.
\end{abstract}

\maketitle

\blfootnote{\hspace{-0.4em}*Please cite the journal version of this paper published in~IEEE Security \& Privacy Magazine 2024.}

\chapteri{I}mage-based sexual abuse~(IBSA), such as the sharing of non-consensual intimate images~(NCII), more colloquially known as~\enquote{revenge porn}, is an increasing problem and has recently gained more legal recognition as a criminal offence, e.g., in the~UK under Section~188 of the Online Safety Act~2023. This issue becomes even more serious when the victims are children, understood as anyone aged under~18 as per the~UN Convention on the Rights of the~Child. Reports from the National Center for Missing and Exploited Children~(NCMEC), one of the largest organisations involved in combating child sexual abuse material~(CSAM) online, show an exponential increase in reported images in recent years\footnote{\url{missingkids.org/gethelpnow/cybertipline/cybertiplinedata}\label{cybertip}}. 
Image exploitation and sexual image crimes are considered serious violations of children's rights and the continuum of harm in these forms of violence and abuse impact many throughout their lifetime~\cite{doi:10.1177/2372732220941534}.
Regardless, such crimes often introduce differential vulnerabilities, i.e., different populations face different types and degrees of security, privacy, and safety risks with serious consequences to their lives.

The challenge of detecting such material has long been an important area of work for both law enforcement and content-based social media platforms like~Facebook, Instagram, and~TikTok. One of the key methods is assessing uploaded content for matches to known problematic content. This is done by creating large databases storing~\enquote{digital fingerprints} of identified content~(rather than the content itself) and checking any images/videos against this during the upload process to prevent distribution.

In the case of~CSAM, the databases of fingerprints are collated and maintained by~NCMEC and similar organisations, then shared with partnering platforms or so-called~\enquote{Electronic Service Providers}~(ESPs). If a match is detected, the~ESP must report this back to~NCMEC, who coordinates with law enforcement to take action against perpetrators. The~NCMEC also runs the online service~\enquote{Take it Down}~(TID\footnote{\url{takeitdown.ncmec.org/}\label{TIDfn}}) that allows users to flag material that shows them underage.

At their core, such tools rely on matching~\emph{perceptual hashes}, which are supposed to be irreversible digital fingerprints of images.
However, it has been shown that some perceptual hash functions, which are also used in other contexts, are prone to attacks that result in avoiding detection of illicit material, causing detection of targeted innocent images, and leaking information about the original image from the hash~(see, e.g., \cite{Dolhansky2020adversarial,jain2022adversarial,struppek2022learning,Bhatia2022exploiting,prokos2023squint}).

In this paper, via an attack covering four prominent perceptual hash functions, we demonstrate that inversion attacks based on generative adversarial networks~(GANs) are feasible with consumer-grade hardware. This includes a study of~Facebook's~PDQ hash\footnote{\url{github.com/facebook/ThreatExchange/blob/main/hashing/hashing.pdf}\label{PDQ}}, which is used by~TID, and Apple's NeuralHash\footnote{\url{apple.com/child-safety/pdf/CSAM_Detection_Technical_Summary.pdf}\label{Apple_psi}}, for which previously \REV{no attacks were known to reconstruct pre-images from hash values}.

\pagebreak

Consequently, perceptual hash values that are uploaded by users and distributed in the clear should be treated almost as carefully as the original images. This is because, for example, a malicious service provider or attacker obtaining such hash values by compromising the service provider's infrastructure could apply our attack methodology to obtain a good approximation of the original sensitive images.

As we find that not only one particular but all studied perceptual hash functions are vulnerable, we argue that the current approach to the design and implementation of services such as~TID is inherently insufficient.
Given the highly sensitive nature of such images/videos and the vulnerable user groups of of such systems, more effort towards data protection is required. We discuss our findings and propose alternative approaches using cryptographic protocols for privacy-preserving~IBSA content removal.

\section{RELATED WORK}

Many online platforms (e.g., Facebook and Microsoft Bing) allow users to manually report inappropriate and~(potentially) illegal content shown to them to the service providers for them to review and remove.
This \REV{can be time-consuming and re-traumatising for victim-survivors who have to find and report the material separately on each different platform, and is wholly insufficient in cases where threatened users want to preemptively report material without knowing whether or where it appears on a platform. 
Recent work by Qin et al.~\cite{qin2024did} studied the technological and interpersonal behaviours adopted by users sharing intimate content to proactively protect themselves from IBSA. They found that users are concerned about the ability for digital fingerprints of their images to be linked back to them, and called for further research. Hence, in this work, we investigate the security of technology for proactive user-led reporting.
However, it should be noted that \emph{user-led} reporting stands apart from the majority of related technology to \emph{automatically} detect~CSAM material either via server- or client-side scanning, which we discuss in this section.} 

\subsection{Server-Side Scanning}
Traditionally, perceptual hash matching is done automatically by online platforms in the background; for this, they compare the hashes of all content stored on and/or passing through their servers with those in curated databases of known~CSAM provided by, e.g., NCMEC. 
Any matches detected by the platform are dealt with via their own policies regarding reviewing the content to ensure it violates their terms of use and then removing it, as well as following legal requirements around reporting content to the relevant authorities.
However, the increasing adoption of~\emph{end-to-end encryption} in both secure messaging apps~(e.g., WhatsApp and~Signal) and file storage~(e.g., iCloud) means that servers are increasingly unable to scan data due to encryption.
As a result, there is an urge to bypass end-to-end encryption by moving to client-side scanning, which we discuss next.

\subsection{Client-Side Scanning}
In~August~2021, Apple~(in cooperation with~NCMEC) announced a~CSAM scanning system for images that are uploaded to~iCloud.
Notably, the proposed system\footref{Apple_psi} would utilize a private set intersection~(PSI) mechanism to detect known~CSAM images on users' devices while still providing end-to-end encryption. 

However, due to severe backlash from privacy advocacy groups\footnote{\url{wired.com/story/apple-csam-scanning-heat-initiative-letter}\label{newman2023apple}}, these plans were ultimately abandoned. 
The raised concerns mainly revolved around the integrity of the encrypted hash database: Both~NCMEC and~Apple could feed in arbitrary hashes if compromised or pressured by governments, \REV{potentially targeting} certain individuals or communities.
\REV{Research has shown that an adversary could even hide a dual purpose such as facial recognition within client-side scanning~(CSS) algorithms~\cite{Jain2023deep}.}
\REV{Although recent work addresses concerns around~CSS through public verification of private hash matching systems~\cite{Scheffler2023public},} Apple in~August~2023 justified abandoning their proposed system \REV{as }~\enquote{it was not practically possible to implement without ultimately imperiling the security and privacy of our users}\footref{newman2023apple}.

Despite this, the~UK Online Safety Bill~2023 includes a provision commonly known as the~\enquote{spy clause}, which requires online platforms to implement scanning of encrypted messages for content associated with child abuse. 
\REV{Whilst the~UK government has conceded that this is currently not~\enquote{technically feasible} and will not be enforced, these powers are still enshrined within the legislation and may be enforced in the future\footnote{\url{ft.com/content/770e58b1-a299-4b7b-a129-bded8649a43b}}.}

\section{PERCEPTUAL HASH \\ FUNCTIONS}
\label{sec:phf}

\emph{Perceptual hash functions}~(PHFs) aim to mimic the human perception of images by giving the same hash value for images that are semantically the same, i.e., hold the same meaning, despite changes such as rotation, cropping, file format, or converting to greyscale.
This contrasts with~\emph{cryptographic hash functions}, where any small change to the original image~(the \emph{pre-image}) will completely change the resulting hash value~(the~\emph{digest})~\cite{Farid2021overview}.
The measure of the extent to which a perceptual hash function's values remain unchanged under such minor image edits is called its \emph{robustness}~\cite{drmic2017evaluating}.
We say two images~$x, y$ are~\emph{semantically similar} if for some distance metric~$d$ and a threshold~$\epsilon$ we have~$d(x,y)<\epsilon$~\cite{Bhatia2022exploiting}.
We define the~\emph{perceptual difference} of two images~$x,y$ as the difference between their~$k$-bit perceptual hash digests, usually using the~\emph{normalized~Hamming distance}:
\begin{equation}
\delta(x,y)=\frac{1}{k}\sum_{i=1}^k|h_i(x)-h_i(y)|.
\end{equation} 
Hence, images~$x,y$ have~\emph{perceptual similarity} if~$\delta(x,y)<\Delta_k$ for some threshold constant~$\Delta_k$~\cite{prokos2023squint}.

In addition to robustness, two other key properties of hashes are distinctness and non-reversibility~\cite{Farid2021overview}.
For a~\emph{distinct}~PHF~$H$ and semantically different~$x$ and~$y$ with respect to some threshold~$\epsilon$ such that~$d(x,y)>\epsilon$, we expect that their digests are different~(i.e., $H(x)\neq H(y)$).
In other words, we expect~\emph{collision resistance}.
For a~\emph{non-reversible}~PHF, it should not be possible to reconstruct the pre-image~$x$ given only the digest~$H(x)$.

\subsection{Attacks on Perceptual Hashes}

Prior research~\cite{prokos2023squint,hooda2022re-purposing} determines that perceptual hashes can be undermined in four methods of attack: 
\begin{enumerate}
    \item \textbf{Detection Avoidance (DA):}
    Modify a harmful image so that it looks almost the same, but the hash is too dissimilar to be flagged.
    Targeting robustness, this type of attack maximizes the difference between perceptual hash digests, yet minimizes the semantic difference.

    \item \textbf{Targeted Second Pre-Image (TSP):}
    Make small imperceptible changes to a harmless image until it has the same hash as a harmful image, or vice versa.
    This can lead to so-called~\emph{poisoning} attacks, where the aim is to generate false-positives through the manipulation of targeted collisions, challenging the collision resistance.

    \item \textbf{Information Extraction (IE):}
    Targeting the non-invertibility property, learn details about the original image from the hash digest, for example, predicting attributes or classes. 

    \item \textbf{Hash Inversion (HI):}
    Learn a mapping between the source images and hashes to generate an approximation of the pre-image if given just the hash value.
    This also contradicts non-invertibility, and is the type of attack we present in this article.
\end{enumerate}

In the following, we summarize previous attacks on perceptual hash functions, \REV{highlighting} a research gap and \REV{addressing it by} presenting the first known hash inversion attacks on~PDQ\footnote{\url{github.com/faustomorales/pdqhash-python}\label{pdq-python}} and~NeuralHash\footnote{\url{github.com/AsuharietYgvar/AppleNeuralHash2ONNX}\label{NH}}, as well as successful attacks against~aHash\footnote{\url{github.com/JohannesBuchner/imagehash}\label{ahash}} and~PhotoDNA\footnote{\url{github.com/jankais3r/pyPhotoDNA}\label{pypDNA}}. 
 
\begin{table*}[t]
    \centering
    \small
	\resizebox{\textwidth}{!}{%
    \rowcolors{3}{gray!25}{white}
    \newcommand\ccg[1]{\cellcolor{white}{#1}}
	\begin{tabular}{lC{1.7cm}C{1.7cm}C{1.7cm}C{1.9cm}|C{0.45cm}C{0.6cm}C{0.45cm}C{0.45cm}}
		\toprule
		 & \multicolumn{4}{c}{\textbf{Attacked Perceptual Hash Functions}} & \multicolumn{4}{c}{\textbf{Type(s) of Attack}}	\\\hhline{~--------}
		\ccg{\textbf{Reference}} & \ccg{\textbf{aHash} \footref{ahash}}             & \ccg{\textbf{PDQ} \footref{PDQ}}   & \ccg{\textbf{NeuralH.} \footref{Apple_psi}}        & \ccg{\textbf{PhotoDNA} \footref{microsoft}} & \ccg{\textbf{DA}}                & \ccg{\textbf{TSP}}        & \ccg{\textbf{IE}}         & \ccg{\textbf{HI}}         \\
		\midrule
		Locascio 2018 \footref{Locascio2018}   & \checkmark        &           &          &          &                 & \checkmark &          & \checkmark \\
		Dolhansky \& Canton-Ferrer 2020~\cite{Dolhansky2020adversarial} &  \checkmark       &         &           &           &                  & \checkmark &           &           \\
		Athalye 2021 \footref{athalye} &                 &  &           &  \checkmark         &                  &           &           & \checkmark \\
		Jain et al.~2022~\cite{jain2022adversarial} & \checkmark        &  \checkmark         &  &           & \checkmark        &           &           &           \\
		Struppek et al.~2022~\cite{struppek2022learning} &                  &           & \checkmark          &  & \checkmark        & \checkmark & \checkmark &          \\
		Bhatia \& Meng 2022~\cite{Bhatia2022exploiting}&                  &           & \checkmark          &  & \checkmark        & \checkmark & \checkmark &           \\
		Prokos et al.~2023~\cite{prokos2023squint}&                  & \checkmark &  & \checkmark          & \checkmark        & \checkmark &           &       \\
		\textbf{This work }&       \textbf{\checkmark}           & \textbf{\checkmark} & \textbf{\checkmark} & \textbf{\checkmark}          &        &  &           &    \textbf{\checkmark}    \\
		\bottomrule
	\end{tabular}
    }
 \caption{Synthesis of attacks against perceptual hash functions investigated in this work. Studied attacks are detection avoidance~(DA), targeted second pre-image~(TSP), information extraction~(IE), and hash inversion (HI).}
 \label{tab:attacks}
 \vspace{-1em}
\end{table*}

In~2018, Nick Locascio showed that it is possible to reconstruct an approximation of the original image from~aHash values by using the~Pix2Pix conditional GAN\footnote{\url{towardsdatascience.com/black-box-attacks-on-perceptual-image-hashes-with-gans-cc1be11f277}\label{Locascio2018}}.
We introduce this methodology in the next section and extend it to attack three prominent perceptual hashes currently used for CSAM detection:~PDQ\footref{PDQ}, NeuralHash\footref{Apple_psi}, and~PhotoDNA\footnote{\url{news.microsoft.com/2009/12/15/new-technology-fights-child-porn-by-tracking-its-photodna/}\label{microsoft}}.
Locascio also performed poisoning attacks on~aHash, modifying the~Pix2Pix objective to~\REV{TSPs}, which would cause false positives.
Work by~Jain et al.~\cite{jain2022adversarial} further demonstrated~\REV{DA} attacks for~aHash.
Attacks on~PhotoDNA followed soon after this proprietary algorithm was leaked\footnote{\url{hackerfactor.com/blog/index.php?/archives/931-PhotoDNA-and-Limitations.html}\label{krawetz}}, with both~\REV{TSP} and~\REV{DA} attacks by~Prokos et al.~\cite{prokos2023squint}, as well as~Ribosome, a~HI attack by~Anish Athalye\footnote{\url{anishathalye.com/inverting-photodna/}\label{athalye}}.
Along with~PhotoDNA, there have been both~\REV{TSP} and~\REV{DA} attacks on~PDQ by~Prokos et al.~\cite{prokos2023squint} and on~NeuralHash by~Struppek et al.~\cite{struppek2022learning}.
For the latter, an~\REV{IE} attack was also realized, correctly classifying up to~52\% of hashes in certain classes and categories, followed by novel black-box attacks on~NeuralHash by~Bhatia and~Meng~\cite{Bhatia2022exploiting}.
We give a concise comparison in~\autoref{tab:attacks}.

\subsection{Case Study: Take It Down}
The website \enquote{Take It Down}\footref{TIDfn} was launched by the National Center for Missing~\& Exploited Children~(NCMEC) to provide a secure and private way for minors and adults to report photos and videos that were taken of them under the age of~18 and involve~\enquote{nude, partially nude or sexually explicit} content.
In particular, the user-selected material is not sent to anyone or uploaded anywhere.
Instead, a perceptual hash of the content is computed locally -- the original material never leaves the reporter's device.
The hash value is then uploaded to a database of user-reported child sexual abuse material~(CSAM) and checked for matches against the hash values of all~(new) content uploaded to partnering Internet platforms.
These platforms include major social media sites like~Facebook, Instagram, and~TikTok, as well as several adult video sites~(e.g., OnlyFans).
\REV{In 2023, the Take It Down site received over 2 million page views, resulting in more than 48,000 reports and 89,000 hashes\footref{cybertip}.}

\begin{figure*}[htb]
    \centering
    \includegraphics[width=0.95\linewidth]{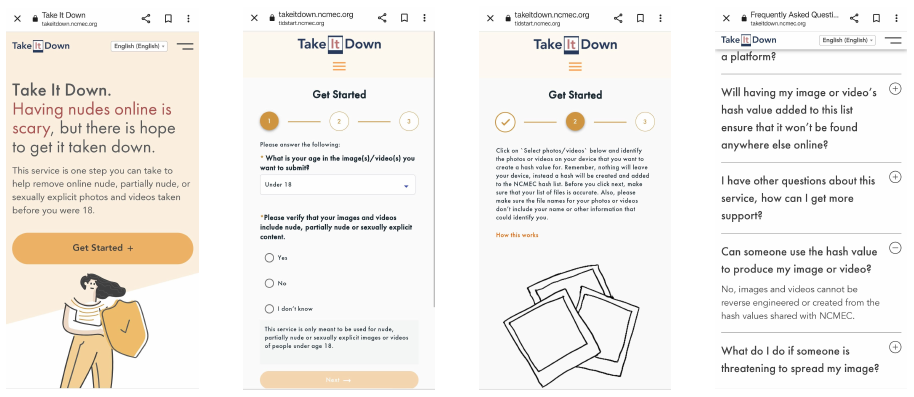}
    \caption{Landing page for the~\enquote{Take It Down} service~(left), and when~\enquote{Under 18} is chosen for age verification~(middle two), and~FAQs regarding the security of hash values~(right).}
    \label{fig:tid}
\end{figure*}

By inspecting this webpage, we identified~\verb|pdq-photo-hasher|~JavaScript and~Web\-Assembly files, verifying that the service is using Facebook's~PDQ perceptual hash algorithm, which was released open-source in~2019, in collaboration with~NCMEC.
Comparing the output of our~PDQ implementation provided an exact match with the visible portion of the hash value displayed by the~TID website.
The site claims to users that any material they report \enquote{cannot be reverse engineered or created from the hash values shared}~(cf.~\autoref{fig:tid}), which we \REV{more closely} investigate in our work.

\section{PERCEPTUAL HASH \\ INVERSION ATTACKS}
We now present our hash inversion attack against aHash, PDQ, NeuralHash, and PhotoDNA.
For this, we describe the methodology and machine learning pipeline used as well as the attack environment and our experimental results.
Finally, we discuss ethical considerations and limitations of our work.

The aim of the attack is to train a model to generate an image as close to the source image as possible, given only the hash value of the source image.
More technically, for a specified target image~$x$ and perceptual hash function~$H$, our goal is to generate a new image~$\tilde x$ from~$H(x)$ such that the difference between the pre-image~$x$ and our output~$\tilde x$ is minimized.
To do this, we follow \REV{Locascio's approach}\footref{Locascio2018} to train a~Pix2Pix-based~\cite{pix2pix2017} conditional~GAN on pairs~$(H(x_i), x_i)$ of target images~$x_i$ and their corresponding perceptual hash~$H(x_i)$ in \REV{a visual pixel representation}.
After training the model, we may test any perceptual hash to generate an approximate~\enquote{inversion} or reconstruction of the pre-image.

\subsubsection{Dataset}
\label{sec:dataset}

Following the same approach as previous hash inversion attacks\footref{Locascio2018}, we use the~CelebFaces Attributes~(CelebA) dataset\footnote{\url{kaggle.com/datasets/jessicali9530/celeba-dataset/}\label{celebA}}, which contains over~200,000 images of various celebrities. 
This dataset was chosen since not only is it widely-used and publicly available, but it also fits well with our use case, allowing us to show the possibility of reconstructing human features from hashes, without using any explicit or indecent content.
\REV{To demonstrate feasibility with low-budget} off-the-shelf resources, we use a random training subset of size~$n=1000$ images.

\subsubsection{Computing Perceptual Hashes}
\label{sec:computing_hashes}

We select four perceptual hashes to test for our experiment: aHash for comparison with the~Locascio attack\footref{Locascio2018}, PDQ for its use in NCMEC's~TID tool\footref{TIDfn}, NeuralHash for its prominence in the topical discussions on client-side scanning for~CSAM detection\footref{Apple_psi}, and~PhotoDNA as a proprietary~Microsoft technology that is used by the~CyberTipline and offered as part of~Microsoft's Azure cloud platform to scan content uploaded by users\footref{microsoft}.

For the former three algorithms, we represent the~0s and~1s of the binary hashes as a square of black and white pixels, according to the corresponding binary value. Since the~PhotoDNA hashes contain instead values between~0-255, we set each pixel in the grid with corresponding~0-255 greyscale values rather than purely black or white.

\subsubsection{Generative Adversarial Network Pipeline}
\label{sec:gan_pipeline}

Pix2Pix~\cite{pix2pix2017} is a conditional~generative adversarial network~(GAN) designed for~\emph{image-to-image translation} in a broad variety of tasks, for example, turning maps into aerial images\REV{,} or sketches into photo-realistic images.
The idea of~\enquote{translation} between images can be thought of very similarly to translation between languages~\cite{pix2pix2017}:  
Two different words in two different languages may represent the same thing.
Likewise, there are many different ways of representing the same image.
For example, AI image generation prompts often include phrases such as~\enquote{$\dots$ in the style of~Van Gogh}, in which case the objective is to translate one image to another image that represents the same thing but in a different~\emph{style}. 
Analogously, the pairs of pre-images and perceptual hash digests represent the same concept in different ways, and our goal is to translate between them, using a~Pix2Pix-style~GAN.

\begin{figure*}[htb]
    \centering
    \scalebox{1.35}{%
	\includegraphics[trim={0 0 0.5cm 0},clip]{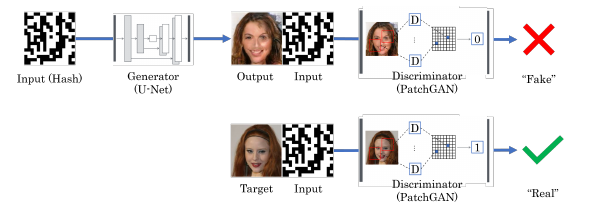}
    }
    \caption{Pix2Pix GAN architecture. Given an input hash, the~U-Net style generator creates fake images, while the~PatchGAN discriminator attempts to determine which image pairs are real or fake.}
    \label{fig:p2p-GAN}
\end{figure*}

GANs consist of two neural networks or~\enquote{adversaries}: the~\emph{generator} and the~\emph{discriminator}.
The generator~$G$ learns to generate fake images to fool the discriminator, whilst the discriminator~$D$ learns to differentiate between real and fake images.
In a standard~GAN, the generator learns only from random noise, whereas in a~\emph{conditional GAN} some additional conditioning information is given to both the generator and discriminator~\cite{pix2pix2017}.
In this case an input image -- the hash -- as illustrated in~\autoref{fig:p2p-GAN}.

\vfill\null

\begin{figure}[htb]
    \centering
    \resizebox{\columnwidth}{!}{%
	\includegraphics[]{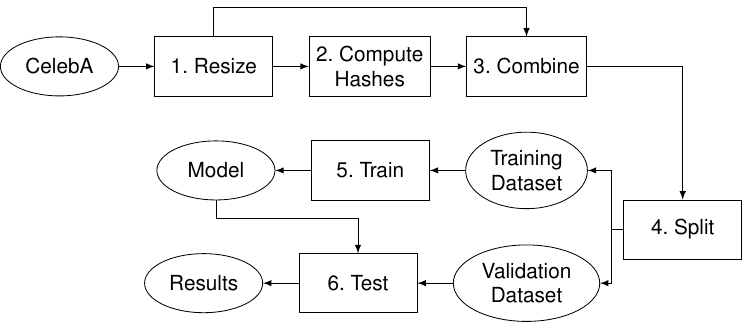}
    }
    \caption{Training and testing pipeline for hash inversion attacks using Pix2Pix-style GAN.}
    \label{fig:pipeline}
\end{figure}

Before training, it is necessary to undertake some preprocessing on the~CelebA dataset, as shown in~\autoref{fig:pipeline}.
First, the images are resized to~256x256 pixels.
Then, they are combined side-by-side into one image with the computed perceptual image hashes, to be parsed by the training model.
The final preprocessing step is to split the combined images into two datasets for training and valuation.
We then created the model from the training dataset by using the~\verb|pix2pix.py| script for~500 epochs, taking around~86 hours using limited resources as described below.
This model was then run on our validation dataset to produce the results shown in~\autoref{1000test500}.

\subsubsection{Training and Evaluation Setup}

After initial proof-of-concept experiments during development using the free resources of~Google Colab, all subsequent training and testing was run using a~Tesla~M60 GPU with~8GB~memory.
Released in~2015, the~Tesla~M60 demonstrates a low barrier to entry for executing this attack and can be considered consumer-grade hardware nowadays, with only~4096~NVIDIA CUDA cores and~1178~MHz boost clock speed -- compared to, for example, a current~GeForce~RTX~4090 with~16384 cores and~2.52~GHz boost clock speed.

\section{Results}

\begin{figure*}[t]
    \centering
    \resizebox{\linewidth}{!}{%
	\includegraphics[trim={0 0 0.5cm 0},clip]{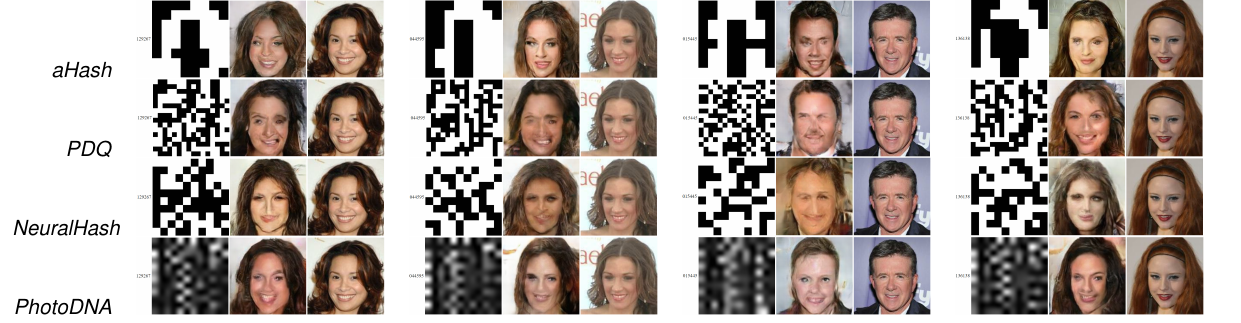}
    }
    \caption{\REV{Example Results: Each triplet shows (left to right) the visual hash representation, the inverted image generated by our attack model, and the target image; from (top to bottom rows)~aHash, PDQ, NeuralHash, and~PhotoDNA. All hash inversion attacks used~Pix2Pix~GAN models trained on~1000 images for~500 epochs.}}
    \label{1000test500}
\end{figure*}

The results in~\autoref{1000test500} are striking for all four~PHFs -- keeping in mind the limited time and hardware resources spent training the~GANs.
Whilst not a perfect match to the target pre-image, the test outputs clearly display many identifiable features such as hair color, hair length, face shape and other facial features, as well as some information about the background.

Looking at our examples in~\autoref{1000test500} and subjectively assessing both similarity to the target image and realism, the best results are arguably for~PhotoDNA, followed by either~aHash or~PDQ. 
PhotoDNA has the longest hash values of the four algorithms, which could suggest a greater amount information~(entropy) for the model to learn from to improve accuracy.
Attacks on the simpler~aHash algorithm seem particularly good at reconstructing sharper boundaries both around and within the faces, hence appearing closer to the target image, yet at the cost of a flatter or more~2D image output.
PDQ attacks on the other hand excel at~3D shading compared to the results from the other~PHFs, generating smoother or more realistic faces, but tend to be further from the target image. 

NeuralHash is a~\emph{deep} perceptual hash, based on a trained neural network model, which may mean that the~GAN cannot take as much advantage of the benefits of the pixel-wise loss function as the hash value is not correlated solely to position in the same way as the others.
Nevertheless, these results show that the perceptual hashes used and proposed for use in~CSAM detection are vulnerable to hash inversion attacks, hence the hashes should be considered sensitive material in the same way as the original images.

To evaluate our results more objectively over the entire dataset, we define perceptual difference~(PD) as the normalized~Hamming difference between the perceptual hashes of a target image and attack-generated image, and define perceptual similarity as the inverse percentage~$(1-PD) * 100$.
Using this definition, we can use multiple different perceptual hash functions as evaluation metrics, for example, we can evaluate the perceptual similarity of an image reconstructed from a~PDQ hash value to the original image using~aHash.
However, we do not use~PhotoDNA as an evaluation metric due to the difficulty of comparing the~Hamming differences of binary versus~0-255 valued arrays. 

In~\autoref{tab:ps}, we evaluate both the mean perceptual similarity and the maximum or~\enquote{best-case} perceptual similarity~(from the perspective of an attacker).
Reconstructed images from~aHash and~PhotoDNA hashes have the highest perceptual similarity on average.
Particularly noteworthy are the best-case results, with the~aHash and~PhotoDNA attacks reconstructing pre-images up to~100\% perceptual similarity, and the~NeuralHash and~PDQ attacks reconstructing pre-images up to~96.88\% perceptual similarity, according to the evaluation with~aHash.
We notice that~PDQ as an evaluation metric seems to be particularly good at telling apart original and reconstructed images with similarity scores up to~20\% points lower than when using other functions as evaluation metric.

\begin{table*}[tb]
    \centering
	\scalebox{0.9}{
	\begin{tabular}{@{\extracolsep{4pt}}lR{1.0cm}R{1.0cm}R{1.2cm}R{1.0cm}R{1.2cm}R{1.0cm}R{1.2cm}R{1.0cm}@{}}
		\toprule
		\multirow{2}{*}{Attacked hash} & \multicolumn{3}{c}{Mean perceptual similarity under} && \multicolumn{3}{c}{Max. perceptual similarity under} \\\cline{2-5}\cline{6-9}
		& aHash & PDQ & NeuralH. & \textbf{\O} & aHash & PDQ & NeuralH. & \textbf{\O} \\
		\midrule
		aHash & 96.40\% & 57.70\% & 68.06\% & \textbf{75.05}\% & 100.00\% & 71.88\% & 88.00\% & \textbf{86.63\%} \\
		PDQ & 65.59\% & 54.80\% & 59.90\% & \textbf{60.10\%} & 96.88\% & 69.53\% & 86.00\% & \textbf{84.14\%} \\
		NeuralHash & 71.42\% & 53.18\% & 72.63\%& \textbf{65.74\%} & 96.88\% & 67.19\% & 90.00\% & \textbf{84.69\%} \\
		PhotoDNA & 86.96\% & 61.26\% & 73.89\% & \textbf{74.04\%} & 100.00\% & 72.66\% & 93.00\% & \textbf{88.55\%} \\
		\bottomrule
	\end{tabular}
	}
    \caption{Mean and maximum perceptual similarity for the results of attacks on~aHash, PDQ, NeuralHash, and~PhotoDNA as computed with~aHash, PDQ, and~NeuralHash. Results averaged are highlighted in bold.}
    \label{tab:ps}
\end{table*}

\begin{figure}[htb]
	\centering
    \resizebox{\columnwidth}{!}{%
	\includegraphics[]{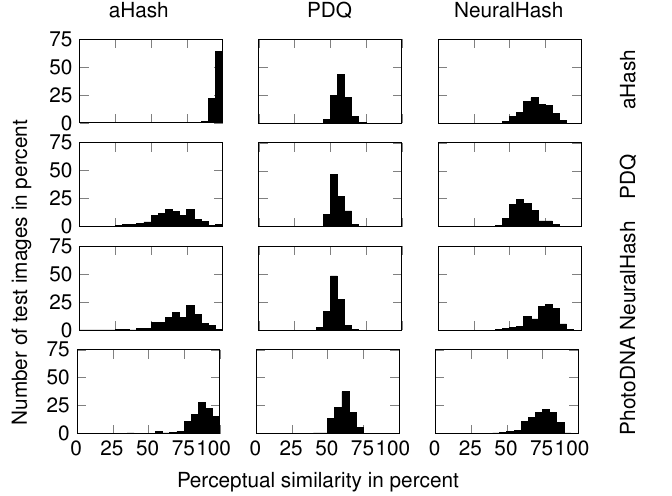}
    }
	\caption{Histograms of perceptual similarity for pre-images reconstructed from~aHash, PDQ, NeuralHash, and~PhotoDNA hashes~(rows) as evaluated using~aHash, PDQ, and~NeuralHash~(columns).}
	\label{fig:eval_graphs}
\end{figure}

For completeness, in~\autoref{fig:eval_graphs}, we also display the frequency distribution of perceptual similarity across all hash attacks and hash evaluation metrics as histograms.
It is evident that the choice of perceptual hash for the evaluation metric has a large impact, however, we do notice some trends independent of this, e.g., aHash and~PhotoDNA consistently attest a higher perceptual similarity.
We note that since the space of all perceptual hash values is much smaller than the space of all images, each matching bit of the hash represents a matching~\emph{feature} of the image.
Hence, even if some distributions shown here reveal that in the worst case we achieve a perceptual similarity of only around~50\%, this still means that we correctly reconstruct~50\% of~\emph{image features}.

\subsection{Ethical Considerations}
\label{sec:ethics}

We ran all attacks solely on hashes of images stemming from a public data set\footref{celebA} that is widely used within the~AI community for evaluation purposes\footref{Locascio2018}.
Likewise, we performed model training only on images of the referenced public data set.

Our results include a study of the~PhotoDNA perceptual hash function.
This algorithm developed by~Microsoft and~Dartmouth College in~2009\footref{microsoft} is proprietary.
However, investigations as early as~2014 uncovered a likely description of~PhotoDNA, and in~2021 a binary implementation of~PhotoDNA was leaked\footnote{\url{https://www.hackerfactor.com/blog/index.php?/archives/931-PhotoDNA-and-Limitations.html}}.
Following previous scientific works studying this hash function~\cite{prokos2023squint}, we obtained its code by extracting it from the publicly available binary installers of forensic toolkits\footref{pypDNA}.

This work received ethical approval from Royal Holloway University.
Following our common practice for responsible disclosure, we informed~NCMEC about the results of our study and how they directly contradict statements in the service's~FAQ regarding the privacy provided when submitting perceptual hash values.
We also offered to share details regarding our proposed secure alternative.
We wrote to them twice (between August and Dec 2023), however, so far, we did not receive any reply and the website did not change.

Making the code and trained models of this paper public might create misuse opportunities for actors with malicious intent.
Therefore, we will not make these artifacts publicly available and only share them with other researchers and organizations such as~NCMEC to help improve such services.
Sharing will only be done after carefully verifying the nature of the request.

\subsection{Limitations}
\label{sec:limitations}

Our experiments were performed solely on images of celebrity faces.
However, existing studies show that generative~AI performs equally well on shapes of entire human bodies~(e.g., \cite{DBLP:conf/cvpr/FruhstuckSSMWL22}).
We leave this as future work.

The training phase of our attack was \REV{severely restricted} by the computational resources and time available. \REV{However, utilising low-budget hardware highlights the ease of executing our attack in real-world scenarios}.
Greater computational power would enable us to train models on complete datasets rather than just small subsets, and for more epochs, resulting in \REV{even closer reconstructions}.
 
We chose to use the distance between perceptual hash values as the similarity metric to objectively measure the quality of our results since these functions are inherently designed to mimic the human perception and assign similar values to similar images.
However, there are many different metrics for measuring the similarity between two images, for example, the~L1 distance per pixel~(as used to train our model), the~L2 norm~\cite{prokos2023squint,Dolhansky2020adversarial}), and the structural similarity or~SSIM~(\cite{Dolhansky2020adversarial,struppek2022learning,Bhatia2022exploiting}).
There also exist~\emph{deep perceptual similarity}~(DPS) metrics, including~AlexNet, ResNet, and EfficientNet~\cite{Dolhansky2020adversarial}.
Determining which (combination of) metrics is best suited to objectively measure the quality of hash inversion attacks is again left as future work.
\REV{Replacing the~L1 distance in our training process with advanced~(DPS) metrics could also improve our attack.}

\section{Secure Hash Matching via \\ Private Set Intersection}
\label{sec:psi}

Our experimental evaluation clearly showed that all commonly used~PHFs are vulnerable to \REV{our inversion attack.
Furthermore, note that our attack is a black-box attack that can be adapted for any~(future)~PHF, although the quality of results may vary.}
Therefore, \REV{perceptual} hash values must be treated almost as sensitive as the original input images.
This begs the question how services like~TID could be realized without \REV{clients having} to transfer perceptual hash values to the service provider in the clear.

\subsection{Introduction to Private Set Intersection}

\begin{figure}[htb]
	\includegraphics[width=0.5\textwidth, trim=0.5cm 0.5cm 0.5cm 1.0cm]{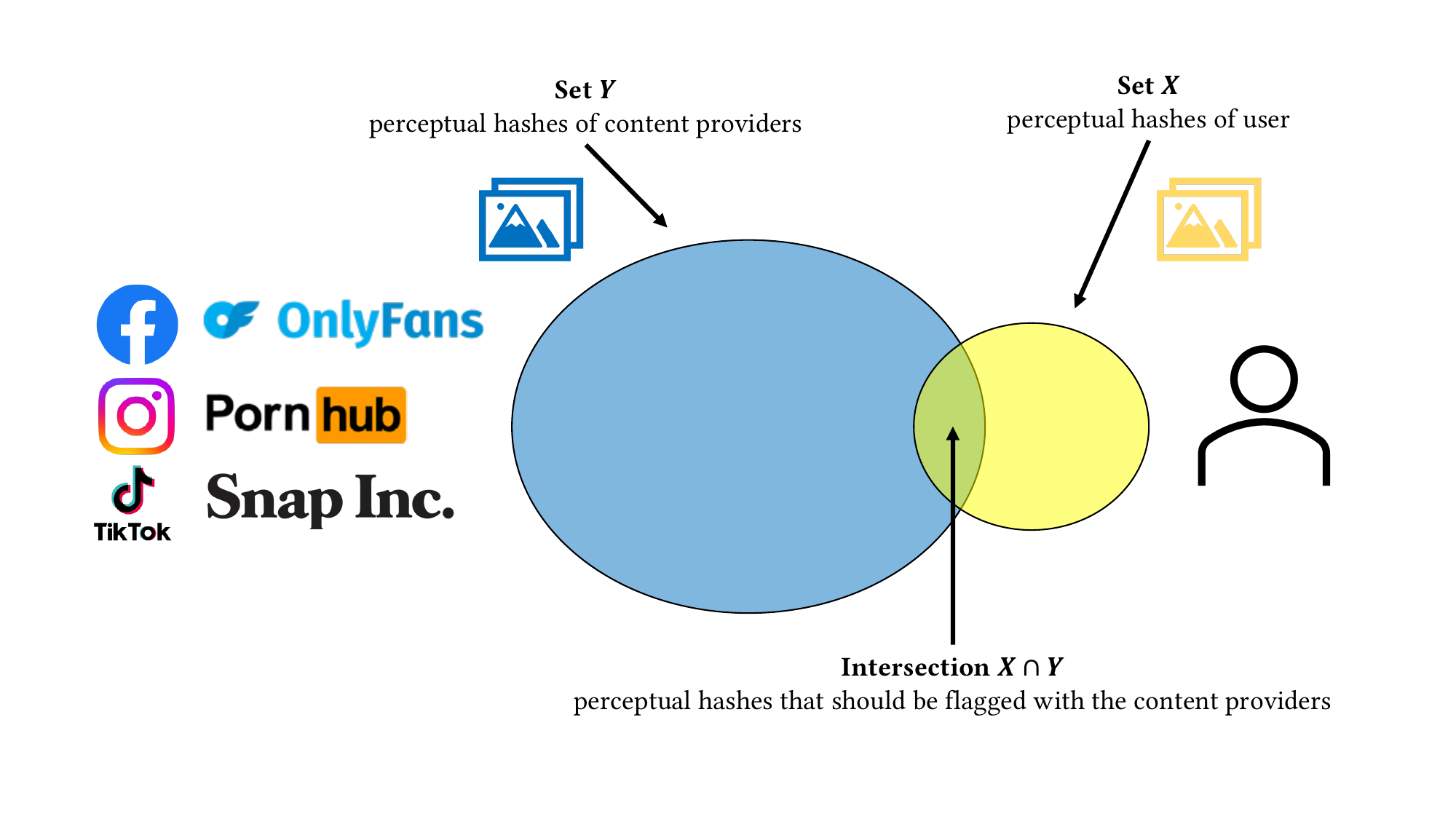}
	\caption{Representing the matching of perceptual hashes between users and affiliated content providers as a set intersection problem.}
    \label{fig:psi_ideal_functionality}
\end{figure}

We propose to map the task of securely matching perceptual hash values to a private set intersection~(PSI) problem~\REV{(cf.~\autoref{fig:psi_ideal_functionality})}:
\REV{The PSI inputs are} a small input set~$X$ provided by the client~(containing the perceptual hash values of images/videos that should be reported) and a large input set~$Y$ provided by the content providers~(containing the perceptual hash values of all images and videos stored at their servers).
\REV{The} goal is to~\REV{reveal only hashes in the intersection}~$X \cap Y$ to the respective content providers.
\REV{This allows the content providers to identify and remove reported material that they store.
In contrast, reported perceptual hashes of unknown material remain confidential.}

\subsection{New \REV{Outsourced} + Unbalanced PSI Variants}

\REV{In the secure computation literature, a wide range of cryptographic~PSI protocols have been proposed~\cite{MORALES2023100567}}.
Unfortunately, \REV{regular two-party} PSI protocols incur a significant computational and/or communication overhead, especially when dealing with large-scale set sizes~(content providers like~Instagram \REV{store} billions of images and videos, i.e., $|Y| > 10^9$).
Luckily, for the use case at hand, an independent third party~(NCMEC) is involved that operates the web service.
\REV{Therefore, we may be able to use optimized~\emph{outsourced PSI} protocols}.
\REV{Furthermore}, since the number of user hashes is much smaller than the number of hashes stored by the content providers, i.e., $|X| << |Y|$, we are in the case of \emph{unbalanced PSI}, \REV{for} which optimized protocols \REV{have been proposed}.
\REV{However, to the best of our knowledge, there exists no~\emph{outsourced unbalanced PSI} protocol.}
As part of future work, we therefore plan to design, implement, and thoroughly evaluate a solution through \REV{combining the advantages of both protocol categories}.

\section{Discussion}

The results of our evaluation highlight that the security and privacy assumptions of tools and services associated with sensitive topics such as~IBSA can be wrong.
In the case of the~\enquote{Take it Down} service, in its~FAQ segment\REV{~(cf.~\autoref{fig:tid})}, there is a question asking:~\enquote{Can someone use the hash value to produce my image or video?} and the answer is:~\enquote{No, images and videos cannot be reverse engineered or created from the hash values shared with NCMEC.}
Our results clearly show that this is fundamentally wrong and the consequences of our attacks can differentially put users at risk. 

One example that is closely linked to our study is the case of sextortion. 
NCMEC's~Cyber Tipline Report from~\REV{2023}\footref{cybertip} reveals that an area with one of the biggest increases in reports was online enticement, up \REV{more than~300\%} from~2021.
Sextortion refers to offenders blackmailing someone for financial or sexual gain, by coercing them to send explicit photos, then threatening to publicly share those images if their target does not send them money or more explicit content.
Historically, offenders tended to coerce girls in order to receive more images, however, in recent years offenders have been increasingly targeting boys \REV{for} financial motives.
In~February~2023, the~FBI issued a joint statement with agencies from across~Canada, UK, Australia, and~New Zealand\footnote{\url{https://www.fbi.gov/news/press-releases/international-law-enforcement-agencies-issue-joint-warning-about-global-financial-sextortion-crisis}}, warning about this~\enquote{global financial sextortion crisis} which has resulted in~\enquote{more than a dozen suicides} related to these events.
Tools such as~TID could offer a way to take the power away from perpetrators, as \REV{preemptively} reporting the images one is being threatened with should prevent them from ever being uploaded.
However, it also highlights the essential high standards of security and privacy required by such a tool, due to the serious consequences of leaking any information about the images they are being entrusted with by vulnerable users of the service.

Besides concerns regarding the security and privacy offered by current implementations of~IBSA content removal tools, the usability and accessibility of such services are of concern too. 
For example, when users are supposed to select their sensitive files for~\enquote{scanning}, a longer description is available under~\enquote{how this works}.
However, this information only remains visible for around~15 seconds at a time -- far too short to read and comprehend the given text.
Particularly, when considering that the target audience includes teenagers and even younger children, it is not clear if the provided explanations of the process are sufficient and intuitive enough for them to completely understand what will happen to their images and videos, and the potential risks to their privacy.
Therefore, we advise to revise the approach to convey this information, and, as part of future work, we plan to explore this design and usability problem together with domain experts and participatory user studies.

Our recommendations for service providers and users are twofold.
First, we suggest that the technology designers consider private set intersection for secure hash matching.
Second, we recommend that users of currently deployed~IBSA reporting and removal tools carefully consider the implications of uploading perceptual hashes.
Given the reality of powerful reconstruction attacks, it might be advisable to only report material of which users are sure that it is already available online or when users have serious reasons to believe the material will soon be uploaded to platforms that are partnering with the respective removal service.
Otherwise, preemptively reporting material that is not yet and potentially never will be published, and thereby exposing its vulnerable hash values, might cause more harm.
We are interested to study user's perspective about such risks, which we leave as future work.

We invite the relevant stakeholders to continue research in such an important research topic, including developing secure, privacy-preserving, usable, and accessible systems.
This requires studying user perspectives and protective actions~(more specifically, with victims and survivors of such crimes), exploring the role of law/policy-making and enforcement as well as education, and eventually offering more inclusive solutions to protect children and other users against image-based sexual abuse material.

\section{CONCLUSION}

People who experience any form of sexual abuse are in vulnerable, high-risk positions and need support to cope with the situation.
We have shown that attacks on perceptual hash functions, which technologically underpin currently deployed~IBSA removal tools, can severely threaten the security and privacy of users engaging with such systems.
Our striking hash inversion results covering a wide-range of prominent perceptual hash functions (including the first known inversion attacks for Apple's NeuralHash and Facebook's PDQ), obtained with very little computational resources, clearly demonstrate that additional protections are necessary when sharing such hash values of sensitive content.
We thus advocate to investigate the deployment of provably secure privacy-preserving technologies such as private set intersection protocols, that would allow for continued matching functionality in the context of~IBSA removal tools without exposing the demonstrably vulnerable perceptual hash values.

\def\refname{REFERENCES}

\end{document}